\documentclass[twocolumn,showpacs,preprintnumbers,amsmath,amssymb]{revtex4}

\newcounter{ctr}

\usepackage{graphicx}
\usepackage{epsfig}
\usepackage{dcolumn}
\usepackage{bm}

\begin{document}
\title{State Differentiation by Transient Truncation
in Coupled Threshold Dynamics}
\author{Yoshinori Watanabe$^1$ and Kunihiko Kaneko$^{1,2}$}
\affiliation{
$^1$Department of Pure and Applied Sciences, University of Tokyo, \\
\normalsize
3-8-1 Komaba, Meguro-ku, Tokyo 153-8902, Japan \\
$^2$ ERATO Complex Systems Biology Project, JST, 3-8-1 Komaba,Meguro-ku,
Tokyo 153-8902, Japan \\ }

\begin{abstract}
Dynamics with a threshold input--output relation commonly exist in
gene, signal-transduction, and neural networks. Coupled dynamical systems of
such threshold elements are investigated, in an effort to find differentiation of elements induced by the interaction.  
Through global diffusive coupling, novel states are found to be
generated that are not the original attractor of single-element threshold
dynamics, but are sustained through the interaction with the elements
located at the original attractor. This stabilization of
the novel state(s) is not related to symmetry breaking, but is
explained as the truncation of transient trajectories to
the original attractor due to the coupling. Single-element dynamics
with winding transient trajectories located at a low-dimensional manifold and having turning points 
are shown to be essential to the generation of such novel state(s) in a coupled system. 
Universality of this mechanism for the novel state generation and its relevance to biological cell 
differentiation are briefly discussed.
\end{abstract}


\maketitle

\section{Introduction}

Differentiation of identical units through interaction is an
important issue both in physics and biology. Through developmental
process, cells with identical genes start to take different chemical
compositions; this process is known as cell differentiation\cite{Cell,Newmantext}. Several
distinct types of cells are generated that take different compositions of
gene expressions. Theoretically, on the other hand, state differentiation of identical
units has been studied in dynamical systems, by using a coupled dynamical
system, such as coupled chaotic systems \cite{GCM} or coupled oscillators \cite{Okuda}.

In coupled dynamical systems with identical elements, there is a homogeneous state 
in which all elements take an identical value. If this
homogeneous state is unstable, differentiation of elements follows. This has been studied
extensively as symmetry breaking. For example, by losing the synchronization in
oscillations, elements are differentiated into clusters with different phases of
oscillation, in coupled chaotic or coupled oscillator systems\cite{GCM,Okuda}. 
Here the differentiation occurs with regards to the phase of oscillation.

In a biological cell, differentiation is more drastic. Different
composition of chemicals or, in other words, different types of gene
expression appear, and the differentiation is not with regard to the
phase of oscillation but in fixed composition of chemicals (e.g. proteins).
To describe cell differentiation as a coupled
system of intra-cellular oscillatory dynamics, isologous
diversification has been proposed\cite{cdiff1,cdiff2}, while its dynamical systems
analysis is not as yet fully developed.

In gene expression dynamics of a cell, the basic process is on/off output against input(s), 
with some threshold
function, rather than oscillatory dynamics\cite{gene-net}. In the present paper, we report a novel
mechanism for state differentiation, by taking elements with such
threshold function (i.e., $y=tanh(\beta x)$ with
$\beta>1$)and coupling them globally with each other through simple
diffusion coupling. By varying several parameters, we find the generation
of novel states in a coupled system that are not attractors of the
original single-element dynamics. This differentiation of states
is not explained as symmetry breaking and, indeed, the original attractor also remains
stable. To explain such coupling-induced
generation of novel stable states, we propose a transient truncation
mechanism, which brings about stabilization of stagnation point(s) in
transient trajectories, mediated by interaction with elements that have already fallen
on the original attractor. We show how this mechanism works, and
describe the condition for the generation and
stabilization of novel states. Generality of this mechanism
in a coupled threshold dynamics model is discussed,
as well as its extension and relevance to cell differentiation.

\section{Model}

Gene expression\cite{gene-net} and signal transduction\cite{signal} as well as neural response\cite{NN} often follow
threshold dynamics, where the output $y$ can be represented by $y=tanh(\beta x)$ as $x$ as an
input and $\beta(>1)$ as a parameter representing the sensitivity.
The input to each variable (gene or neuron) comes from
several genes (or neurons) that are connected
through excitatory or inhibitory couplings.
For such dynamics, the following threshold dynamics model is
often adopted;

\begin{equation}\dot{x}_i=\tanh[\beta\sum_{0\leq j}^{M-1}
J_{ij}x_j-\theta_i]-x_i,
\end{equation}

\noindent
where $x_i$ is the state of $i$th variable (e.g., gene expression)
with $i=0,1,...,M-1$, while the component of connection matrix $J_{ij}$
takes a value from positive to negative values\cite{threshold}. Here, we study
the case in which $J_{ij}$ takes either $-1$ or $1$ that is selected
randomly, as is used in the spin-glass model\cite{SG}, while the specific form of
the distribution of $J_{ij}$ is not important for discussion of the result. The threshold value $\theta_i$ is fixed, and is also
distributed over [-1,1]. The parameter $\beta$, representing the
sensitivity, is fixed at 4 in the present paper, while the behavior to be
discussed is unchanged as long as $\beta >1$. 

Now, we consider an ensemble
of elements, each of which follows the same equation (1) as
single-element dynamics, and introduce interaction among such elements. For example,
consider $N$ cells, each of which has identical gene expression
dynamics. Then, the global behavior of an ensemble of these 
cells is represented by the above intra-cellular dynamics and
interaction among them. Instead of $x_i$, we need to study the
dynamics of the variable $x_i(k)$, the state of the $i$th component
(e.g., $i$th gene) of the $k$-th element (cell). Here we take the
simplest form of interaction, diffusive, global coupling, to
all elements(cells).
Now, the model we discuss is written as

\begin{eqnarray}
\dot{x}_i(k)=\tanh[\beta\sum_{j=0}^{M-1} J_{ij}x_j(k)-\theta_i]- x_i(k)\nonumber\\ 
+D_i (\bar{x}_i-x_i(k))
\label{eq:model_mean}
\end{eqnarray}
with $k=1,2,...,N$, and $\bar{x}_i=(1/N)\sum_{\ell=1}^N x_i(\ell)$ is the average
value of the $i$th component over all elements, while $D_i$ is the
strength of this diffusive coupling over elements. We use the present
mean-field model (global coupling) as an idealized basic system.

Of course, another choice in the coupling form is spatially local
interaction, such as the nearest-neighbor diffusion coupling among
elements located on a lattice. Here, we use the above global interaction, 
because we are interested in the basic property of
coupled threshold dynamics and state differentiation. 
In general, with the choice of spatially local
interaction, differentiation of state values by elements appears more
easily, while the mechanism to be described for the global interaction
works even for the local interaction case.

In a biological context, this type of model was discussed, for instance by
Mjolsness et al.\cite{Mjolsness} and  Salazar-Ciudad  et al.\cite{Sole},
in relationship with the problem of  cell differentiation,
where these authors chose the interaction $J_{ij}$ and (local) cell-cell
interaction to meet a specific biological situation. Here, we are
interested in general features of this class of models, so that we have
chosen the simplest situation, as described above. In a physics context, the
above model (with local coupling) was studied analytically and
numerically by Hansel and Sompolinsky\cite{Somp}, as a model for
spatiotemporal chaos, where their interest is focused on the limit with $M
\rightarrow \infty$, and fully chaotic behavior. Our interest in the
present paper lies in the differentiation into distinct stable states
(mostly fixed points) for a system with a relatively small $M$.

\section{Generation of Novel States by Transient Truncation Mechanism}

The single-element dynamics (1) (or the model (2) with $D_i=0$ for all $i$)
can have multiple attractors in general, which are either
a fixed-point, limit cycle or strange attractor. To discuss the interaction-induced
generation of novel states other than the attractor(s) of a single element
dynamics (1), however, it would be better to study the case with only one
attractor at first.

In fact, the behavior of a coupled dynamical system has been studied
extensively, when an element system has only one limit-cycle or chaotic attractor. If the
attractor is a limit cycle, synchronization among elements often
occurs through the coupling, while, if the attractor is chaotic,
clustering of elements into several states can occur\cite{GCM,Okuda}. In the latter case,
state values are differentiated by elements, as a result of the
instability of the homogeneous (synchronized) state, while the
differentiation is understood as symmetry breaking. Indeed, in our
model (2), such clustering is generally observed when the single-element
dynamics (1) shows chaotic or oscillatory dynamics.

On the other hand, if the attractor of single-element dynamics (1) is
a fixed point, a homogeneous state of the fixed point over all
elements is always stable in the present diffusive coupling system.
Then, the generation of novel states other
than the fixed point is not possible by the
symmetry-breaking mechanism. However, we have 
found several examples in which
the coupled system (2) exhibits differentiation of state values,
when started from initial conditions far from a homogeneous state.
Inhomogeneous states with $x_i(k)\neq x_i(j)$ are observed,
for some network $J_{ij}$, and for some values of $\{D_i \}$, and $\theta(i)$.

\begin{figure}[htbp]
\begin{center}
\includegraphics[scale=0.85]{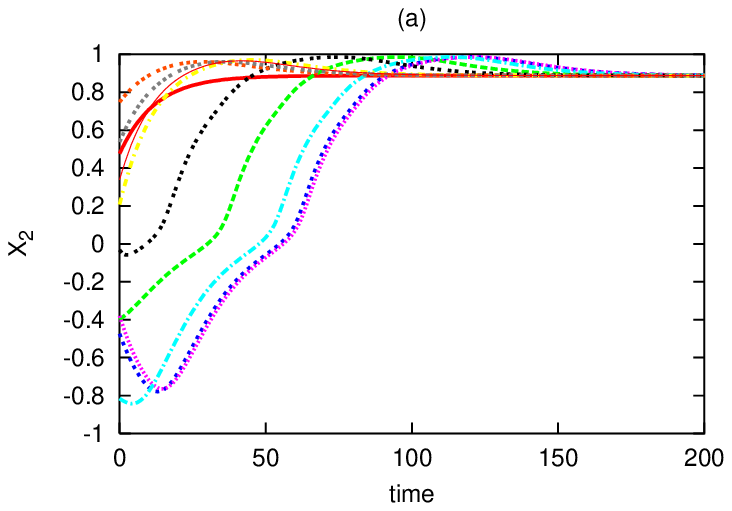}
\includegraphics[scale=.85]{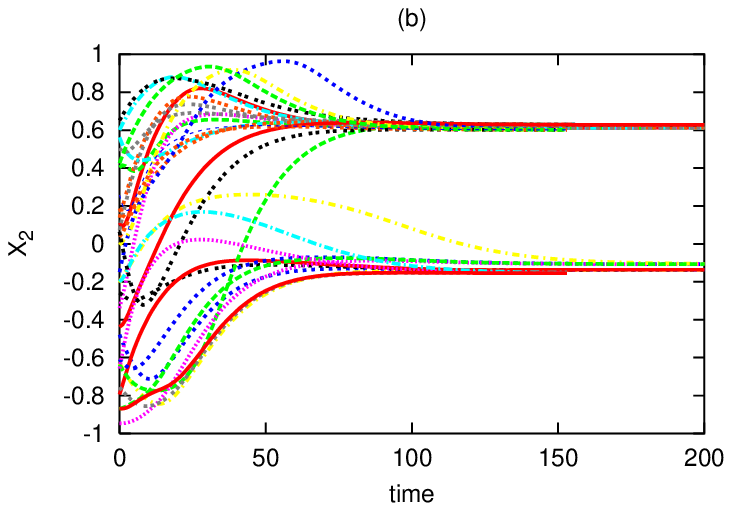}
\caption{(Color Online) Time series of $x_2$ of the threshed dynamics model with $M=5$.  (a) 
Overlaid time series of single-element dynamics (1), from 10 initial conditions chosen 
randomly.  From all initial conditions, a single, fixed-point attractor is reached.
(b) Time series of the coupled dynamics model (2) overlaid for 30 elements 
chosen from $N=100$ elements,
for a single initial condition.  Other than the original fixed point of (a), another
fixed-point state is reached, which is stabilized by the interaction. The matrix $J$ and
parameter values are chosen as follows.}
\label{fig:5_17j}
\end{center}
\end{figure}

\begin{figure}[htbp]
\begin{center}
\includegraphics[scale=0.85]{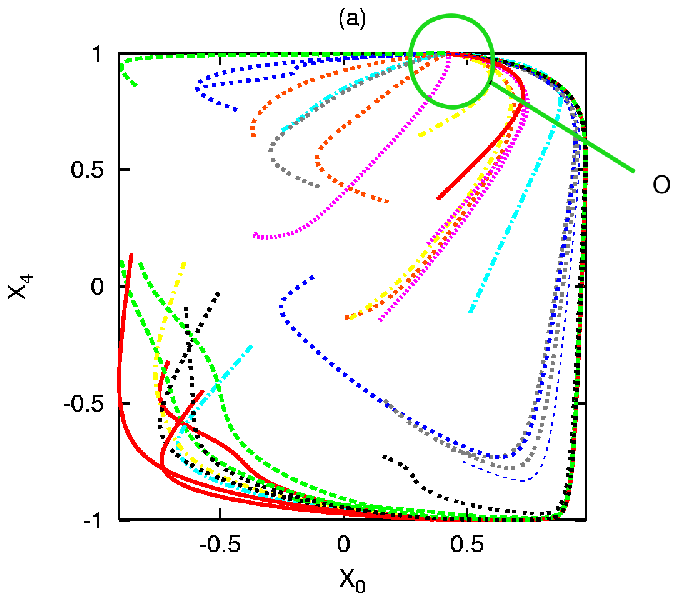}
\includegraphics[scale=.85]{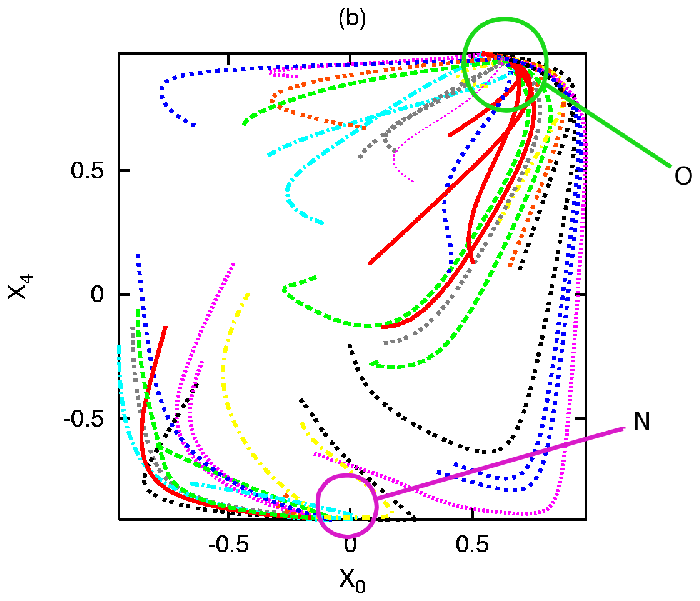}
\caption{(Color Online) The trajectory of the model corresponding to Fig.1.  $(x_0(k),x_4(k))$ 
is plotted as a projection to two-dimensional plane from the 5-dimensional phase
space. (a) Single-element dynamics corresponding to Fig.1(a). Overlaid plot over 25
initial conditions.  All the orbits are attracted to the fixed point denoted by "O".
(b) Coupled dynamics model corresponding to Fig.1(b). Overlaid
plot for 30 elements chosen from N=100.  The orbits are either attracted to the original attractor shown
as "O" (a green circle), or to
a novel fixed point state denoted by 'N' (represented by a violet circle).  The latter is
stabilized by the interaction .  About  one third of elements are
attracted to 'N', while others are attracted to the original attractor 'O'.
Note that the location of the state corresponding to the
original attractor is slightly shifted by the coupling term with the elements
at the novel state.}
\label{fig:5_17}
\end{center}
\end{figure}

An example of such behavior is shown in  Fig.\ref{fig:5_17j} and Fig.\ref{fig:5_17}, 
where $M=5$.  The corresponding single element dynamics (1) has only a fixed-point
attractor. Here we choose
\begin{eqnarray}
J&=&\left(
\begin{array}{ccccc}
+ & - & - & + & +  \\
- & - & - & + & +  \\
+ & - & - & - & +  \\
+ & + & - & + & +  \\
- & + & - & + & +  \nonumber
\end{array}
\right)\\
\end{eqnarray}
where $+$ denotes 1 and $-$ denotes -1, 
$\theta=$(0.14, -0.75, 0.71, -0.78, 0.32), and
$D=$ (0.95, 0.027, 0.30, 0.18, 0.95).
while several other choices of $J$, $D_i$, and
$\theta(i)$ give rise to the similar behavior. The time series of $x_2$ of a single-element dynamics
(1) are plotted in Fig.\ref{fig:5_17j}(a) by taking a variety of initial conditions,
which shows the relaxation to a single, fixed-point attractor. In
Fig.\ref{fig:5_17j}(b), the time series of $x_2(k)$ over several
elements are plotted. One can see differentiation of final state values
into two fixed-point values, one of which corresponds to the original fixed-point
value of the single element dynamics (1), although the value of the fixed point is
slightly shifted, due to the coupling term.  The other fixed-point value, on the other hand,
does not have a correspondent in the single-element dynamics.

Corresponding to these time series, we have plotted a two-dimensional projection of orbits
from the 5-dimensional phase space. Fig.\ref{fig:5_17}(a)
again shows the single-element dynamics (1) without interaction. Each line
represents time evolution of $(x_0,x_4)$ starting from different
initial conditions, while Fig.\ref{fig:5_17}(b) shows the evolution of
the coupled system (2), where an orbit from a single initial condition
is plotted, with each line as an orbit of each element. One can again
see clearly that a novel attracting state ('N') other than the original fixed
point attractor $x_i(k)=x_i^*$ ('O')is created through the interaction.

Recall that the homogeneous state with $x_i(k)=x_i^*$ for all elements
$k$ is always stable. Indeed, when initial condition is set so that
the states of all elements are located near this fixed point, the
attractor $x_i(k)=x_i^*$ is always reached. In this sense, the present
mechanism differs distinctly from the clustering or other mechanisms
based on spontaneous symmetry breaking. Besides the stable
homogeneous state, there appears a macroscopic state
consisting both of the elements at the original fixed point and a novel fixed point,
when the initial condition of elements is set far from homogeneity.
(In most examples, we choose
a random initial condition where $x_i(k)$ is taken randomly
from [-1,1].) Here the novel fixed point 'N' is stabilized by the coupling 
with other elements located at 'O'.

\begin{figure}[htbp]
\begin{center}
\includegraphics[scale=1.2]{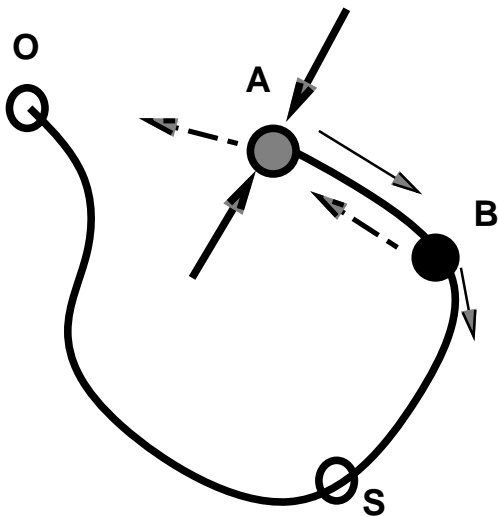}
\caption{(Color Online) Schematic representation of transient-truncation mechanism. Original relaxation
of single element dynamics takes a course $A\rightarrow$ $B\rightarrow$ $S\rightarrow O$,
while the interaction with elements near $O$ hinders the relaxation from $A$ to $S$.}
\label{fig:sche}
\end{center}
\end{figure}

Indeed, by taking a variety of networks $J_{ij}$, we have observed
several examples of formation of such novel state(s), and found a common
mechanism. The mechanism of the generation of novel
state(s) other than the original fixed-point attractor is explained as
follows: 

Consider the case in which a single unit dynamics has long-winding
transient trajectories before they reach the original unique fixed point
('O'), as shown in Fig.\ref{fig:5_17}(a).
During the transient process, the orbit has (a few) turning points at
which the motion of $x_i(t)$ is slowed. While some elements have
reached the original fixed point fast, others are still on a route to
it. At some of turning points, the relaxation of an element
take a course once going farther away from the original fixed point 'O'. 
On the other hand, the diffusive
coupling with elements that has already reached the original final fixed point
drives the transient element toward it ( see Fig.\ref{fig:sche} for
schematic representation). This coupling suppresses the relaxation of
single-element dynamics toward the original fixed point. When the
directions of the original relaxation and the attraction to
the original fixed point are opposite, the two driving forces may
balance each other around a turning point where the motion is
stagnated (see 'stagnation' point 'S' in Fig.\ref{fig:sche}).
Then, the relaxation to
the original fixed point is truncated, and some elements remain around
this stagnation point, to create an interaction-induced novel state,
as shown in Fig.\ref{fig:5_17} and Fig.\ref{fig:sche}.

This 'transient-truncation' mechanism works under the following condition;

(1) The loci of transient orbits are restricted within a
low-dimensional manifold: If the transient trajectories cover a high-dimensional
region in the phase space, orbits from different elements approach the
original  attractor from a variety of directions, and the transient
truncation by the diffusion coupling does not work effectively. 
When the transient-truncation mechanism works, many points
reach the original attractor, taking a specific course restricted within low-dimensional
region in the phase space,
as displayed in Fig.\ref{fig:5_17}. 
Contraction to the low-dimensional manifold is so strong that each element is located
within a low-dimensional manifold, as shown schematically in Fig.\ref{fig:sche} 
(as thick red arrows toward $A$), 

(2) The transient orbit has one or several turning points. At some
turning point, the single-element motion is stagnated where
the orbit stays for a long time, so that the driving force by the single-element
dynamics is weak there.  Hence, the diffusive coupling to the original fixed point
is sufficient to stop the original relaxation course.
See Fig.\ref{fig:5_17} for example.

(3) The direction of transient orbit around this stagnation point is
roughly ``opposite" to the direction to the original fixed point, 
attracted by the diffusive coupling.
Then, the orbit is trapped around this stagnation point, as shown in Fig.\ref{fig:5_17}. 

Since the interruption of transient dynamics is caused by
the diffusive coupling to the elements already fallen on the original fixed point,
the degree of interruption depends on the number of such elements, which is denoted by $N_f$.

\begin{figure}[htbp]
\begin{center}
\includegraphics[scale=1.2]{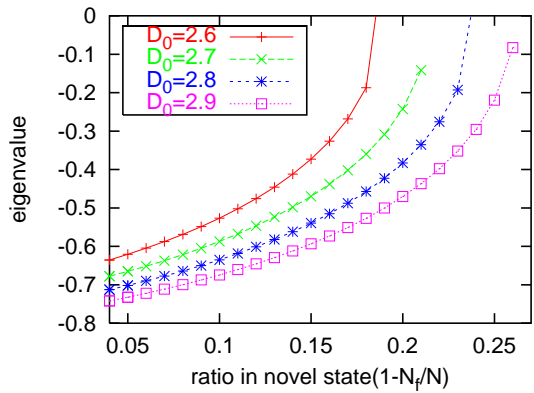}
\caption{(Color Online) Dependence of the stability of the novel state upon $N_f$, the number of elements
located at the original fixed-point attractor.  By using the same model
for Fig.(1)-(2), we have computed the largest eigenvalue of the Jacobi matrix 
of the novel fixed-point state, by fixing the number of elements at the original
attractor at $N_f$.  Here the eigenvalue is plotted as a function of $(N-N_f)/N$,
i.e., the fraction of the number of elements at the novel state.
When the eigenvalue exceeds zero, the state is no longer stabilized, 
and the exponent is not computed.  Different symbols corresponds to 
different sets of diffusion coupling $D_i$, which is changed by fixing
$D_1/D_0=0.448 $, $D_2/D_0=0.19$, $D_3/D_0=0.078 $,$D_4/D_0=0.052$, 
and changing only $D_0$, as shown in the figure.}
\label{fig:kkae}
\end{center}
\end{figure}

To study how the stability of the novel state changes with $N_f$, we
have computed the largest eigenvalue of the Jacobi matrix of the evolution
equation at this novel fixed point ('N') induced by coupling. If the 
eigenvalue is negative, this novel fixed-point state is stable. This eigenvalue
depends on the number $N_f$, (or more generally, on the ratio $N_f/N$).
We have plotted this eigenvalue against the ratio $1-N_f/N$, i.e., the fraction
of the elements at the novel state.
As shown in Fig.\ref{fig:kkae}, the
eigenvalue is negative only if $N_f$ is larger than some threshold,
while it decreases with the increase of $N_f$. In other words, the novel
state is sustained only under the existence of a moderate number of the
elements at the original fixed point.  Existence of the threshold
number for $N_f$ is natural, since the new state is sustained by
''attractive force'' to the original fixed point.

\begin{figure}[htbp]
\begin{center}
\includegraphics[scale=0.9]{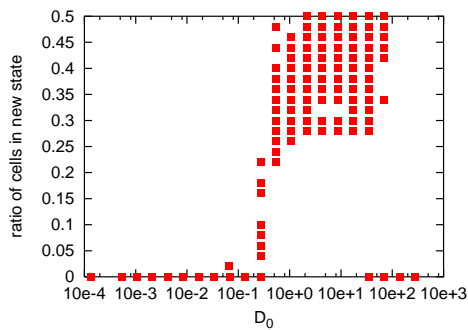}
\caption{(Color Online) Fraction of elements fallen on the novel state
is plotted against the strength of diffusion coupling.
The fraction is measured over 50 randomly chosen initial conditions.
Diffusion strength $D_i$ is changed by
fixing $D_i/D_0$ and varying $D_0$, in the same way
as in Fig.4, and changing only $D_0$ as shown.
The fraction is plotted against $D_0$.  For $D_0<0.1$ and $D_0>100$, the 
novel fixed-point state does not exist, and all the elements fall on the original fixed point.}
\label{fig:5_17d0}
\end{center}
\end{figure}

According to the above mechanism, the appropriate strength of diffusion
coupling is necessary to stabilize the novel state. Indeed, the
present transient truncation mechanism works only for a given range of diffusion
constants. If it is too small, the attraction to the original fixed
point is too weak to interrupt the relaxation course of the single-element
dynamics, so that all the elements fall on the original fixed point.
On the other hand, when the diffusion coupling is too
large, the diffusion coupling dominates so that all the elements take the
same value. Then, the dynamics follow the single-element
dynamics (1), so that all elements fall on the original fixed point.
(See Fig.\ref{fig:5_17d0} for the diffusion constant dependence of the
existence of the novel state, where $D_i$ is changed by keeping the
proportion among  $D_i's$ (i.e., fixing $D_i/D_j$)).

\section{Generalization}

We have studied the behavior of the model (2) by taking
a variety of networks and by changing $N$ and $M$, to
find that the generation of novel states by the mechanism of the
last section is general. 
We have computed the
fraction of the networks that show the generation of novel state(s) 
from fixed-point attractor(s), by the above mechanism.  
The fraction of the network (and $\theta_j$) for such behavior
remains to be $1\sim 5$ \%, when $M$ is changed from 8 to 64.  
Here, we have computed 100 networks for each of $M$, and the 
transient-truncation mechanism from fixed point attractor(s)
is observed for $1\sim5$ networks among them.
We also note that the present mechanism also works, even if
the coupling $J_{ij}$ is sparse, in the sense that many of
$J_{ij}$'s  are set at 0.  For example, we have observed the novel state generation
by transient truncation, with a similar fraction,
for a system with $J_{ij}=0$ for 70 \% of the matrix.

\begin{figure}[htbp]
\begin{center}
\includegraphics[scale=0.8]{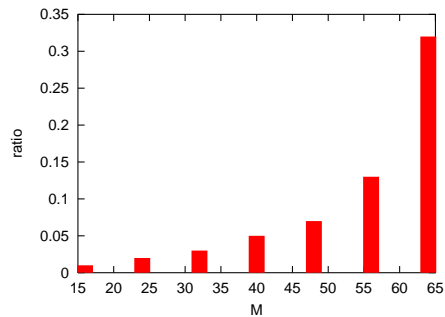}
\caption{(Color Online) Fraction of networks that exhibit oscillatory
dynamics for single-element dynamics.
For each value of $M$, we have chosen 100 networks with randomly chosen $J_{ij}$,
and carried out the simulation without coupling, to check if there is a limit-cycle 
or a chaotic attractor. For the corresponding coupled system,
clustering of elements into a few groups is observed with the
increase of the coupling strength, and  then a synchronized state over elements appears for 
the further increase of coupling, as is studied in globally coupled maps\cite{GCM}. 
$N$ is fixed at 128.}
\label{fig:Mkae}
\end{center}
\end{figure}

Another type of non-trivial
behavior of the coupled system (2) is clustering of elements
into different phases of oscillations, when the single element
shows (chaotic) oscillation, as was discussed
in globally coupled maps\cite{GCM}. The fraction of networks showing
the clustering also increases with $M$, as shown in Fig.\ref{fig:Mkae}.
This is natural, since chaotic behavior is more frequently observed 
in a single-element dynamics\cite{Somp}.
In other words, the network only with fixed-point attractors 
for single-element dynamics decreases with $M$.  Hence, among the  
networks having only fixed-point attractors, the fraction to show the 
transient-truncation mechanism slightly increases with $M$\cite{limitcycle}. 

Although the generation of coupling-induced novel state(s) is
common to the networks above, it is often more complicated
than the simple example in the last section where the original
single-element dynamical system has only a single, fixed-point
attractor.  

\begin{figure}[htbp]
\begin{center}
\includegraphics[scale=0.85]{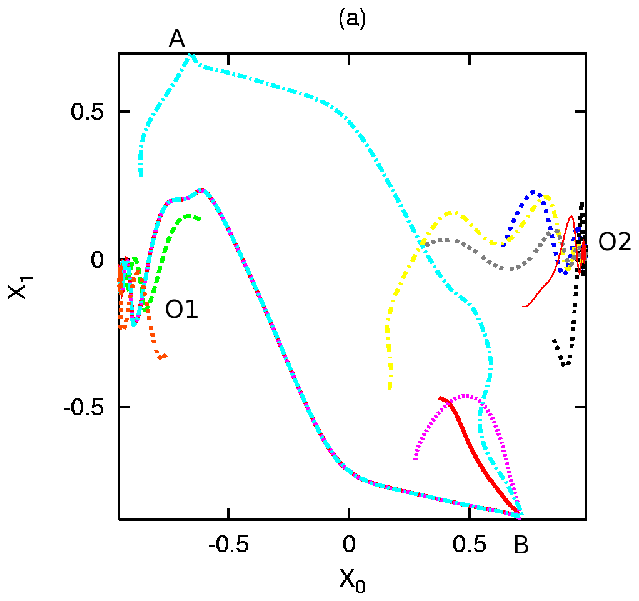}
\includegraphics[scale=0.85]{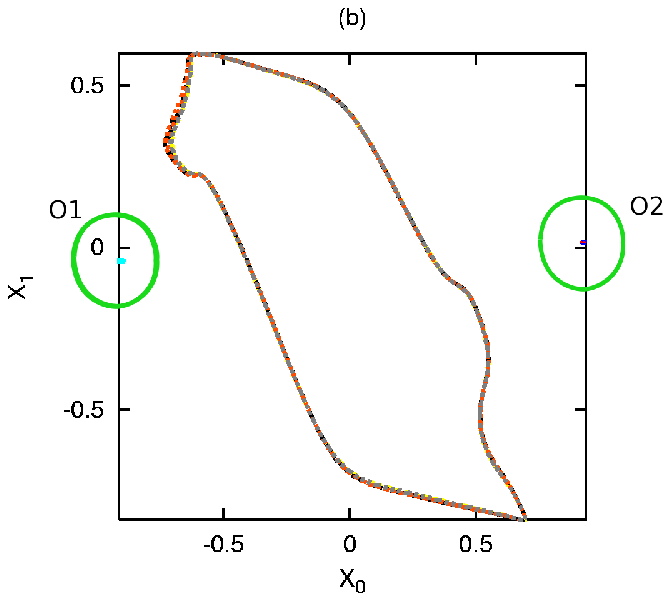}
\includegraphics[scale=0.85]{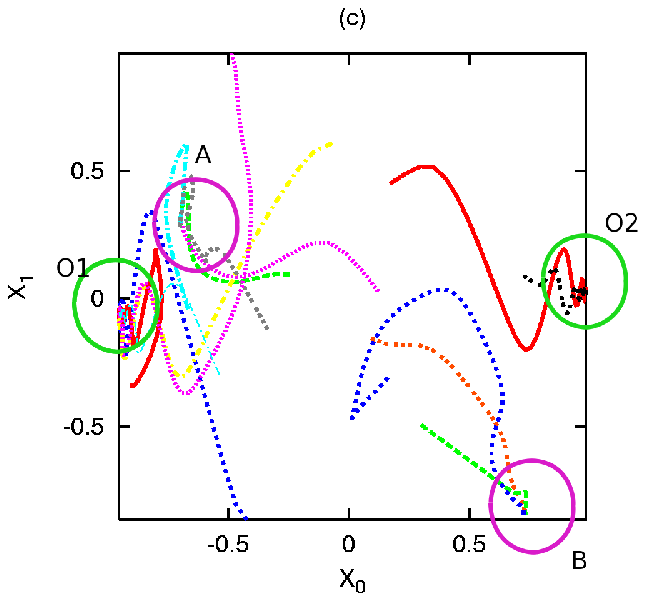}
\caption{(Color Online) The trajectories $(x_0(k)(t),x_4(k)(t))$ 
are plotted, as a projection to a two-dimensional plane from the M-dimensional phase
space ($M=10$). (a) Single-element dynamics (1), overlaid plot over 20
initial conditions.  (b) Coupled dynamics model (2) corresponding to (a), with the coupling
values $D=$ (0.20, 0.27,  0.66,  0.41, 0.26, 0.15, 0.32, 0.29, 0.096, 0.33). 
Final states are plotted for all of 100 elements.  
The original fixed-point attractors are shown as
O1 and O2. At (a), there are winding transient orbits that fall on the
attractor O1 or O2. In (b), a novel limit cycle is generated.
(c) The trajectories of a coupled dynamical system with the same set of
$J_{ij}$ and $\theta(i)$, but by a different set of $D_i$, i.e.,
$D=$ (0, 0.30, 0.0037, 0, 0.79, 0, 0, 0.0070, 0.80, 0.0028).  With the coupling,
novel fixed-point states A and B are generated, besides the original fixed points, instead of the limit cycle in (b). Trajectories of 12 elements among 100 are plotted. The matrix $J$ and parameter values are chosen as follows.
}
\label{fig:10_195d0}
\end{center}
\end{figure}

Even if the single-element dynamics has multiple attractors,
the transient-truncation mechanism still works.
Indeed, when $M$ is large, we observed the case in which novel states are generated by the coupling
besides the original multiple attractors.  For example, when there are two fixed point attractors
in the original dynamics, other fixed-point(s) or a limit cycle state is stabilized
due to the coupling, as a result of transient truncation.
An example is shown in Fig.\ref{fig:10_195d0}, where we choose
\begin{eqnarray}
J&=&\left(
\begin{array}{cccccccccc}
- & - & + & + & + & + & - & - & - & +\\
+ & - & + & - & + & + & + & + & - & +\\
+ & + & - & + & + & + & - & + & - & +\\
+ & + & + & + & + & - & - & - & - & -\\
- & - & - & + & - & + & - & + & + & +\\
- & + & - & + & - & - & + & + & + & -\\
- & - & - & + & + & + & - & + & - & -\\
+ & - & - & - & + & - & - & - & - & +\\
+ & + & - & - & + & + & - & + & - & +\\
- & + & + & - & - & + & - & + & - & -\nonumber
\end{array}
\right)\\
\end{eqnarray}
with $\theta=$( -0.27, 0.98, 0.22, -0.25, -0.92, 
0.63, 0.44, 0.64 0.74, -0.73).

In this example, there are two fixed points, denoted by O1 and O2,
in the original single-element dynamics (1). There are
transient trajectories that have a few turning points,
and that are attracted to O2, as shown in Fig.\ref{fig:10_195d0}(a). With the
coupling to
elements located at O1 and O2, the transient trajectory is truncated, and
a limit cycle is generated for the remaining elements,
as shown in Fig.\ref{fig:10_195d0}(b). This truncation
is possible only if the numbers of elements at O1 and O2 are
within some range, but the range is rather broad, so that
the coupling-induced limit cycle state is observed
just by starting from random initial conditions.
Furthermore, we have observed this type of novel limit-cycle state
in a variety of networks.

Formation of multiple novel states is also possible in some networks
(and with suitable choice of diffusion). In Fig. \ref{fig:10_195d0}(c), we show an example
of formation of two fixed-point states, by using the same network as
in Fig.\ref{fig:10_195d0}(a)(b), but by taking a different set of diffusion couplings
$(D_0,D_1,...,D_{M-1})$ given in the figure caption. Here, two fixed-point states, A and B, are generated
around two stagnation points. By starting from some initial
conditions, both of these two stagnation points become stable due to
the interaction with other elements. 

Here, the novel fixed-point state A exists under the presence of B; otherwise
the elements locating at around A cannot stay there, but move toward B.
Hence there is ordering between A and B. The latter states are
necessary for the former, but not vice versa.
Generally, when there are several stagnation points S1,S2,.., along a low-dimensional
transient orbit, and coupling-induced novel states are formed accordingly
as N1,N2,..., there is
ordering with regards to their existence, as
N1 exits under the presence of N2,N3,..., and N2 exists under N3,N4,.., and so forth.

In some other networks, novel states A and B mutually stabilize each other;
the state A exists under the presence of elements at B, and vice versa.
By removing all elements taking the state A,
elements taking the state B become unstable and are absorbed into the
original fixed point, and removing the elements taking B also results in
the destabilization of the state A.

Finally, the original attractor of the single-element dynamics need not
necessarily be a fixed point. The mechanism of the transient truncation
can work even if the original attractor is not a fixed point, but a
limit cycle, as long as there are stagnation points along the transient
orbits satisfying (1)-(3) in the last section.

\section{Summary and Discussion}

In the present paper we have studied a coupled threshold dynamics
model, to find emergence of novel states stabilized by the
coupling. Although we have adopted just a simple global
diffusive coupling which tends to homogenize all the element
values, there appears
differentiation of the state values, induced by the coupling.
The mechanism of the generation of novel states is explained
as the truncation of transient orbits that are located on a
low-dimensional manifold in the phase space.
The interaction with the elements that have fallen on the original
attractor suppresses the relaxation process of the remaining elements
at some stagnation point, to make it a  novel
fixed point (or a limit cycle).

The transient-truncation mechanism is based just on the existence of
'winding' transient orbits on a low-dimensional manifold, with
several turning points. Hence, the coupling-induced
formation of novel stable states by this mechanism
is not restricted to the present model. It should be
generally possible in coupled dynamical systems, with
the above class of transient orbits at a single-element level.

\begin{figure}[htbp]
\begin{center}
\includegraphics[scale=0.8]{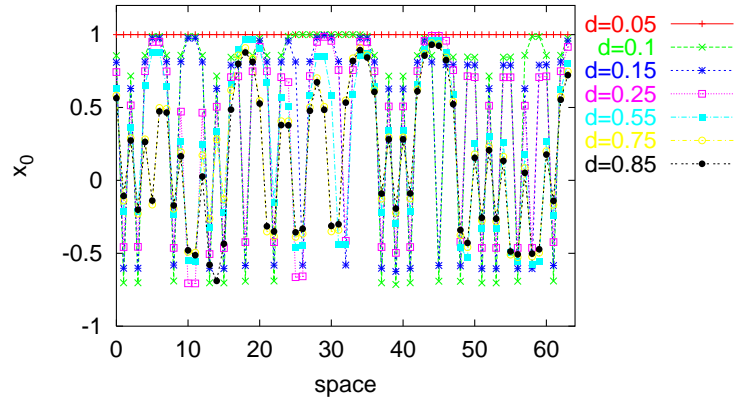}
\caption{(Color Online): Snapshot pattern of $x_0$ for the model (3), with one-dimensional nearest-neighbor
interaction.  $M=10$ and $N=64$ with a periodic boundary condition.
When $d_i$ is less than some value ($\sim 0.07$), there exists only a single attractor with a spatially homogeneous state 
with $x_0 \approx 1$.  On the other hand, for larger $d_i$, spatially inhomogeneous patterns are reached as attractors 
for most initial conditions.}
\label{fig:space}
\end{center}
\end{figure}

Still, we have not found such differentiation in the
previous studies on coupled dynamical systems in which a much simpler
element is adopted(such as the logistic map\cite{GCM}). At least, one can conclude
that the present transient truncation appears more frequently in the coupled
threshold dynamics model given by eq.(2). We expect that this is
because of typical nature of transient orbits in (1).  Indeed, in
threshold dynamics, each variable tends to approach either 
1 or -1. At some points with $x_i \simeq 1$ or -1 for some
$i$'s, the change in the variable values is slowed, and then the
trajectory departs from them. The transient dynamics of a single
element often involve such switchings between +1 and -1 with
stagnation of motion. Thus the requirement on transients discussed in
\S 3 is easier to be satisfied in the present model
than in coupled oscillators or coupled chaos.

Another clear example of such stagnation is a heteroclinic cycle\cite{hetero,Chawanya,Tachikawa}.
Although the heteroclinic cycle itself is not a transient orbit to a
fixed point required here, slight structural perturbation on the
heteroclinic cycle can lead to transient orbits on low-dimensional
manifold with some stagnation points. At this point, it is
interesting to recall that a class of threshold-network dynamics can
generally produce heteroclinic cycles \cite{Rabinovich}.

Of course, generation of novel states is important in the study of
biological cell differentiation. As
the number of cells increases through the developmental process,
they interact with each other, and some cells start to
exhibit different gene-expression patterns. Indeed,
the spontaneous cell differentiation process
has been discussed theoretically as isologous diversification\cite{cdiff1,cdiff2}.

Considering that eq.(1) is a simplified form of
gene-expression dynamics, the present mechanism of state
differentiation may be relevant to the cell differentiation,
since novel states stabilized by the (cell--cell) interaction
take gene-expression patterns distinct from those of
the original attractor. Indeed,
mutual stabilization and hierarchical ordering of cell types, observed
in the present model, may be important to the discussion of
robustness and irreversibility in the cell differentiation process\cite{whatlife}.
Here, it is interesting to note that long
transient dynamics on a low-dimensional manifold has recently been
observed in a gene network model constructed from biological
data\cite{Ouyang}.

To close the paper, we again note that inclusion of spatially local
interaction to the present study is quite straightforward.  In a one-dimensional
lattice, one can adopt a nearest-neighbor diffusive interaction model as given by

\begin{eqnarray}
\dot{x}_i(k)=\tanh[\beta\sum_{j=0}^{M-1} J_{ij}x_j(k)-\theta_i]- x_i(k) \nonumber\\
+d_i (\frac{x_i(k+1)+x_i(k-1)}{2}-x_i(k)).
\end{eqnarray}

In this case, generation of novel states by the present
mechanism works.  Without coupling (i.e.,  by taking $d_i=0$), only a homogeneous
state with a stable fixed point exists, while with coupling spatially inhomogeneous
pattern appears, depending on the initial condition.  "Spots" of novel states are distributed
with some distance, leading to spatial configuration of differentiated
elements.  Note again that this pattern formation is not a result of symmetry breaking
as in Turing pattern\cite{Turing}.

Acknowledgment: The authors would like to thank Koichi Fujimoto,
Masato Tachikawa, Shuji Ishihara, and Satoshi Sawai for stimulating
discussions.


\begin{thebibliography}{999}
\bibitem{limitcycle}
Generation of novel states by the transient-truncation mechanism
can also occur even if the attractor of single-element dynamics
is a limit cycle.  However, we have not systematically counted the fraction
of such cases, since automatic detection for such case is not so easy.
\end{thebibliography}
\end{document}